\documentclass[12pt,preprint]{aastex}
\usepackage{fullpage,rotating,epsfig}

\newcommand{\be}{\begin{equation}}
\newcommand{\ee}{\end{equation}}
\newcommand{\bea}{\begin{eqnarray}}
\newcommand{\eea}{\end{eqnarray}}

\newcommand{\p}{\partial}

\newcommand{\half}{{\textstyle{1\over2}}}
\newcommand{\msun}{\,M_\odot}
\newcommand{\pc}{\,\mbox{pc}}
\newcommand{\mpc}{\,\mbox{mpc}}
\newcommand{\yr}{\,\mbox{y}}
\newcommand{\cm}{\,\mbox{cm}}
\newcommand{\rmpc}{r_{\mathrm{mpc}}}
\newcommand{\au}{\,\mbox{AU}}

\newcommand{\vI}{{\bf I}}
\newcommand{\va}{{\bf a}}

\newcommand{\vb}{{\bf b}}
\newcommand{\vx}{{\bf x}}

\newcommand{\vS}{{\bf S}}

\newcommand{\vOmega}{{\bf \Omega}}
\newcommand{\vphihat}{\mbox{\boldmath $\hat\phi$}}

\newcommand{\vw}{{\bf w}}
\newcommand{\vl}{{\bf l}}
\newcommand{\vv}{{\bf v}}
\newcommand{\vp}{{\bf p}}
\newcommand{\ve}{{\bf e}}
\newcommand{\vz}{{\bf z}}
\newcommand{\vrr}{{\bf r}}

\newfont{\caps}{cmcsc10}
\newcommand{\df}{{\caps df}}
\newcommand{\dfs}{{\caps df}s}

\begin{document}

\title{Secular stability and instability in stellar systems surrounding
massive objects}
\author{Scott Tremaine}
\affil{Princeton University Observatory, Peyton Hall, Princeton,
NJ~08544-1001, USA} 
\email{tremaine@astro.princeton.edu}

\begin{abstract}
\noindent
We examine the stability of a low-mass stellar system surrounding a massive
central object. Examples of such systems include the centers of galaxies or
star clusters containing a massive black hole, and the Oort comet cloud. If
the self-gravity of the stellar system is the dominant non-Keplerian force,
such systems may be subject to slowly growing (secular) lopsided
instabilities. Stability to secular modes is largely determined by the
dependence of the distribution function $F$ on angular momentum $J$. If $\p
F/\p J<0$ at constant energy, all spherical systems are secularly stable. If
$\p F/\p J>0$, as is expected if there is a loss cone at low angular momentum,
all spherical systems in which $F=0$ at $J=0$ (an empty loss cone) are only
neutrally stable, and flattened, non-rotating systems are generally
unstable. These results suggest that secular instabilities may dominate the
structure and evolution of the stellar systems in the centers of galaxies.

\end{abstract}


\section{Introduction}

\noindent
There is strong evidence that most early-type galaxies contain massive black
holes at their centers \citep{kor04}, and tantalizing evidence that smaller
black holes may be present in some globular clusters \citep{vdm03}. These
findings motivate the study of stellar systems in which the gravitational
force is dominated by a central point mass. We call these ``near-Keplerian''
stellar systems.  In this paper we focus on spherical near-Keplerian systems
since these are a natural first approximation for the centers of elliptical
galaxies, spiral bulges, and globular clusters.

Two important features shared by many near-Keplerian systems are: (1) Over a
large range of radii, the dominant non-Keplerian force arises from the
self-gravity of the stellar system. (2) Stars on orbits with near-zero angular
momentum are destroyed---tidally disrupted or swallowed whole---by the black
hole. In combination with two-body relaxation due to stellar encounters, this
process leads to an equilibrium stellar phase-space density or distribution
function (hereafter \df) that is an increasing function of angular momentum,
$\p F/\p J>0$. If the rate at which relaxation or large-scale torques
replenish the lost stars is slow enough, there will be almost no stars on low
angular-momentum orbits (an ``empty loss cone'';
\citealt{ls77}).

A quite different near-Keplerian system is the Oort comet cloud, which
surrounds the Sun at distances of $\sim10^4\au$. The Oort cloud shares the two
characteristic features mentioned above: (1) The mass of the cloud, though
quite uncertain, is large enough that over much of the cloud self-gravity
dominates the other important non-Keplerian forces: the quadrupole field from
the planets on the inside, and the tidal field from the Galactic disk on the
outside (see \S\ref{sec:oort} for quantitative estimates). (2) Comets on low
angular-momentum radial orbits are lost from the Oort cloud by gravitational
interactions with the giant planets.

The goal of this paper is to investigate whether near-Keplerian stellar
systems are subject to slow or secular instabilities. These instabilities
can arise because orbits in Keplerian potentials do not precess, so that
even the weak self-gravity of a low-mass stellar system can systematically
modify the collective orientation of eccentric orbits. The existence and
nature of secular instabilities is independent of the mass $M_\ast$ of the
stellar system relative to the black-hole mass $M$, so long as $M_\ast\ll M$
and self-gravity is the dominant non-Keplerian force; only the growth rate
of any instability depends on $M_\ast$.

\subsection{Timescales}

\label{sec:times}
\noindent
We first review the relevant timescales for the evolution of spherical
near-Keplerian stellar systems. Assume that the system is composed of
$N=M_\ast/m$ stars having mass $m$ and typical orbital radius $r$. The
dynamical time is $t_{\rm dyn}\sim (r^3/GM)^{1/2}$. If the dominant
non-Keplerian force is due to the self-gravity of the stellar system,
then the orbits precess on the secular timescale
\be
t_{\rm sec}\sim t_{\rm dyn}{M\over M_\ast}\sim t_{\rm dyn}{M\over Nm}.
\ee

The two-body relaxation time is \citep{bt87}
\be
t_{\rm relax}\sim {r^{3/2}M^{3/2}\over G^{1/2}mM_\ast} \sim t_{\rm
dyn}{M^2\over mM_\ast}\sim t_{\rm dyn}{M^2\over Nm^2}.
\ee
The energies of the stars diffuse on a timescale $t_{\rm relax}$.

Because the Keplerian ellipse traced out by a star is approximately fixed on a
timescale $t_{\rm sec}$, the average torque between two stars is approximately
constant over this timescale. This slow variation in the mutual torques
between stars gives rise to enhanced or resonant relaxation of the angular
momenta \citep{rau96}; the corresponding timescale is
\be
t_{\rm res}\sim {r^{3/2}M^{1/2}\over G^{1/2}m} \sim t_{\rm dyn}{M\over m}.
\ee
For near-Keplerian systems ($M_\ast\ll M$, $N\gg1$) these four timescales are
well-separated:
\be
t_{\rm dyn}\ll t_{\rm sec}\ll t_{\rm res} \ll t_{\rm relax}.
\ee

Most investigations of near-Keplerian systems assume that they are dynamically
stable. In this case, evolution is driven by relaxation on the timescale
$t_{\rm res}$ or $t_{\rm relax}$. The basis for this assumption is that
$M_\ast\ll M$ so that the self-gravity of the stellar system is small and
collective effects should be negligible. However, the small self-gravity from
the stellar system is responsible for the precession of the orbits and hence
there can be instabilities due to collective interactions that depend on the
distribution of the orientations of the orbits. Such instabilities should
occur on the secular timescale $t_{\rm sec}$ and hence the growth rate would
be faster than resonant or two-body relaxation, no matter how small the mass
of the stellar system may be.

\subsection{Relation between secular, Jeans, and radial-orbit instabilities}

\label{sec:analog}

\noindent
The nature of secular instabilities can be illustrated by a heuristic
comparison to two other instabilities that affect stellar systems: the Jeans
instability that is present in infinite homogeneous systems at wavelengths
that exceed the Jeans wavelength $\lambda_J$, and the radial-orbit instability
that appears in spherical systems with a preponderance of nearly radial
orbits.

Consider an infinite, homogeneous stellar system. We shall consider
perturbations that are independent of the spatial coordinates $y$ and $z$, so
the response of the system depends only on its equilibrium \df\ $F(p)$, where
$p$ is the momentum in the $x$-direction. We shall assume that $F$ is an even
function of the momentum. As usual we invoke the Jeans swindle
\citep{bt87} in which we neglect the contribution of the equilibrium density
to the gravitational field, so the equilibrium Hamiltonian is $H_0(p,x)=\half
p^2/m_i$ where $m_i$ is the inertial mass. We now subject the system to a weak
gravitational potential $\Phi(x,t)=\widetilde \Phi(t)\exp(ikx)$.  The
Hamiltonian is modified to $H(p,x)=H_0(p,x) + m_g\Phi(x,t)$, where $m_g$ is
the gravitational mass (usually $m_i=m_g$, but we shall use the greater
generality). The resulting perturbation to the \df, $f(x,p,t)=\widetilde
f(p,t)\exp(ikx)$, is governed by the linearized collisionless Boltzmann
equation,
\be
{\p f\over\p t}+ \{f,H_0\} +\{F,m_g\Phi\}={\p f\over\p t}+ {\p
f\over\p x}{\p H_0\over\p p} - m_g{\p F\over\p p}{\p \Phi\over\p x}=0,
\label{eq:vlasov}
\ee
where $\{\cdot,\cdot\}$ is the Poisson bracket. In the present context, this
becomes 
\be
{\p \widetilde f\over \p t}+ {ikp\over m_i}\widetilde f - ikm_g\widetilde 
\Phi{dF\over dp}=0. 
\ee
The perturbed density is $\rho(x,t)=\widetilde\rho(t)\exp(ikx)$ where
$\widetilde\rho=\int dp\,\widetilde f$, and Poisson's equation
reads $-k^2\widetilde\Phi=4\pi G\widetilde\rho$.
A neutral mode is present if
\be
k^2=-4\pi Gm_im_g\int_{-\infty}^\infty {dp\over p}{dF\over dp}=
-8\pi Gm_im_g\int_0^\infty {dp\over p}{dF\over dp}\equiv k_J^2.
\label{eq:neut}
\ee
For the usual situation in which $m_i=m_g$, a neutral mode is present if the
integral is negative, which in turn occurs if the \df\ is a decreasing
function of $p$. More detailed analysis shows that the neutral mode divides
stable modes with wavelengths $\lambda < \lambda_J=2\pi/k_J$ from unstable
modes with $\lambda>\lambda_J$ (the Jeans instability). Thus the Jeans
instability is present if $dF/d|p|<0$.

This analysis can be applied by analogy to axisymmetric stellar systems. In
this analogy, the linear momentum $p$ corresponds to the $z$-component of
angular momentum $J_z$, and the position $x$ corresponds to the coordinate
conjugate to the angular momentum, which is the azimuthal angle of the
apocenter, $\omega$. Since $\dot x=p/m_i$, the quantity corresponding to the
inertial mass $m_i$ is an effective moment of inertia $I_{\rm
eff}=J_z/\dot\omega$; since the precession rate $\dot\omega$ depends on the
energy $E$ and angular momentum $J_z$ of the orbit, the effective moment of
inertia is generally also a function of $E$ and $J_z$, which can be positive
or negative.

First consider an axisymmetric stellar system without a central mass point or
strong central density concentration. In this system, nearly radial orbits
cross straight through the center of the galaxy, so that each near-radial
orbit may be considered to have two apocenters separated by
$\Delta\phi\simeq\pi$ radians. In terms of the radial and azimuthal
frequencies $\kappa$ and $\Omega$, each apocenter precesses at a rate
$\dot\omega=\pm(\Omega-\half\kappa)$ where the two signs apply to prograde
($J_z>0$) and retrograde ($J_z<0$) orbits. The ratio $\Omega/\kappa$ is near
$\half$ for radial orbits whose pericenter lies within the central core of the
potential, and $\half<\Omega/\kappa<1$ for orbits with pericenter outside the
core. Thus in general $\dot\omega>0$ for prograde orbits, so $I_{\rm
eff}=J_z/\dot\omega$ is positive; similar arguments show that $I_{\rm eff}$ is
also positive for retrograde orbits. Since the gravitational mass $m_g>0$,
equation (\ref{eq:neut}) suggests that neutral and unstable modes can be
present if $dF/d|J_z|<0$. This is the well-known radial-orbit instability that
arises in galaxies with radially anisotropic \dfs\ (see \citealt{pal94}\ for a
review).

Finally we consider a near-Keplerian system. As we suggested at the beginning
of the paper, in many such systems we expect the equilibrium \df\ to be an
increasing function of angular momentum, $dF/d|J_z|>0$. Equation
(\ref{eq:neut}) implies that an instability may be present when the product
$I_{\rm eff}m_g dF/d|J_z|$ is negative. Thus we expect that an instability may
be present if $I_{\rm eff}<0$, which occurs if prograde orbits have
$\dot\omega<0$. This is often true in disk systems and always true in
spherical systems (see \S\ref{sec:first}). Thus near-Keplerian stellar systems
may be susceptible to an instability analogous to the Jeans or radial-orbit
instability.

Although the Jeans instability is always present in a homogeneous stellar
system, these arguments do not prove that axisymmetric systems with
$dF/d|J_z|>0$ are always unstable. The reason is that in an infinite
homogeneous system there can be arbitrarily small wavenumbers $|k|$, while in
an axisymmetric system the analog to the wavenumber $|k|$ is the azimuthal
wavenumber $m$, which is restricted to integer values.  Modes with $m=1$ are
the most likely to be unstable, but proving stability or instability requires
more analysis, a task we undertake in
\S\S\ref{sec:linpert}--\ref{sec:kep}. 

\subsection{Two examples of near-Keplerian systems}

\noindent
To help relate our theoretical discussion to real systems, we briefly review
the relevant parameters of two familiar near-Keplerian systems, the Galactic
center and the Oort cloud. Our main goal is to establish that there is a
significant radial range in which (1) the potential is near-Keplerian; (2) the
dominant non-Keplerian force is self-gravity, and (3) the loss cone is empty. 

\subsubsection{The Galactic center}

\label{sec:mw}

\noindent
We may describe the mass distribution near the center of the Galaxy by the sum
of a point mass $M=3\times10^6\msun$ (the black hole) plus an approximately
spherical stellar system with density
\be
\rho_\ast(r)=Ar^{-\alpha}.
\label{eq:rho}
\ee
Here $\alpha=1.8$ and the normalization $A$ can be obtained by setting the
enclosed stellar mass $M_\ast(r)=4\pi\int_0^r dr\,r^2\rho_\ast(r)$ to
$1.4\times10^6\msun$ at $r=1\pc$ \citep{Schodel03}. With these parameters,
$M_\ast(r)=M$ at $r=1.9\pc$ (50 arcsec at 8 kpc) so the stellar system is
near-Keplerian for $r\la1\pc$.

We shall measure radii in milliparsecs or \mpc\ (0.026 arcsec at 8 kpc) since
this is not far from the accuracy of the best near-infrared positional
measurements. For circular orbits, the orbital frequency is given by
\be
\Omega^{-1}=0.27\rmpc^{3/2}\,\yr,
\ee
which provides a measure of the dynamical time $t_{\rm dyn}$. The 
apsidal precession rate for nearly circular orbits due to the mass $M_\ast(r)$
is given by
\be
|\dot\omega|^{-1}=3.8\times10^3\rmpc^{0.3}\,\yr,
\ee
and this provides a measure of the secular time $t_{\rm sec}$. The timescale
for resonant relaxation is
\be
t_{\rm res}=t_{\rm dyn}{M_\ast(r)\over m}=100\rmpc^{2.7}\,\yr
\ee
assuming a stellar mass $m=1\msun$. The two-body relaxation time is
\be
t_{\rm relax}=3.5\times10^8\rmpc^{0.3}\,\yr,
\ee
using equation (8-71) of \cite{bt87}, assuming a Coulomb logarithm 
$\log\Lambda=\log(M/m)=15$. 

Solar-type stars are disrupted by the black hole if their pericenter distance
is $<9\times10^{12}\cm=0.003\mpc$ (eq.\ [1] of \citealt{mt99}). The loss cone
is empty for solar-type stars for $r\la 1\pc$ (based on formulae in \S3 of
\citealt{mt99}).

The general-relativistic precession rate for nearly circular orbits is given
by equation (\ref{eq:gr}), 
\be
|\dot\omega_{\rm GR}|^{-1}=6.3\times10^2\,\rmpc^{5/2}\,\yr.
\ee
Thus the self-gravity of the stellar system dominates the precession for $r\ga
2\mpc$.

These simple estimates suggest that the three conditions given at the start of
this section are approximately satisfied in the radius interval $3\mpc\la r \la
1\pc$ ($0.08\arcsec\la r\la 25\arcsec$), a range of over two orders of
magnitude.

\subsubsection{The Oort cloud} 

\label{sec:oort}

\noindent
The outer radius of the Oort comet cloud is determined by the ejection of
comets by gravitational perturbations from passing stars, and is probably
roughly $5\times10^4\au$ in semi-major axis. The inner radius is much less
certain, since Oort-cloud comets with semi-major axes $\la 2\times 10^4\au$
(the ``inner'' Oort cloud) are not normally visible from Earth, for reasons
described below. Models of the formation and evolution of the Oort cloud
\citep{dqt88,ldd01} suggest that the inner cloud may extend to
semi-major axes of $\sim 10^3\au$ or even less, with a mass that may exceed
the mass of the observable outer cloud by a large factor. The total Oort cloud
mass of $\sim 50M_\oplus$ estimated by \cite{wei91} is probably as good as
any, but the uncertainty is at least a factor of five. The recent discovery of
Sedna \citep{bro04} suggests that the inner Oort cloud may contain more mass
and extend to smaller radii than usually believed.

For our purposes it is sufficient to assume a power-law density distribution of
the form (\ref{eq:rho}); we shall choose $\alpha=2.5$ and normalize so that
the mass inside $5\times10^4\au$ is $50M_\oplus$, although steeper
exponents are possible (e.g. $\alpha=3.5$ is estimated by \citealt{dqt88}). 

The orbital frequency for circular orbits is given by
\be
\Omega^{-1}=1.6\times 10^5 r_4^{3/2}\,\yr,
\ee
where $r_4=r/10^4\au$. The apsidal precession rate for nearly circular orbits
due to the mass of the cloud is given by
\be
|\dot\omega|^{-1}=9.4\times10^9r_4\,\yr.
\label{eq:secc}
\ee
The energy and angular-momentum relaxation is dominated by interactions with
passing stars, and the relaxation time is roughly \citep{dqt88}
\be
t_{\rm relax}=1\times10^9 r_4^{-3/2}\,\yr;
\label{eq:relaxc}
\ee
since the dominant interactions are with unbound perturbers there is no analog
of resonant relaxation. 

Comets are removed from the Oort cloud by gravitational interactions with the
giant planets, which either eject them from the solar system or perturb them
into much more tightly bound orbits, if their perihelion distance is $\la
15\au$. The loss cone due to the giant planets is empty if the cometary
semi-major axis is $\la 2\times10^4\au$. Since the Earth is located near the
center of this loss cone, Oort-cloud comets are only visible if they arrive
from the region where the loss cone is full, that is, with original semi-major
axes $\ga 2\times10^4\au$. Thus the properties of the Oort cloud inside this
radius are not directly accessible to observation.

The precession rate of a comet on a nearly circular orbit in the ecliptic due
to the quadrupole moment from the giant planets is
\be
|\dot\omega_{\rm pl}|^{-1}=1.9\times10^{14} r_4^{7/2}\,\yr.
\ee
Assuming the local density in the Galaxy is $\rho_G=0.10\msun\pc^{-3}$, the
precession rate due to the Galactic tide is given by
\be
|\dot\omega_{\rm Gal}|^{-1}=2.2\times10^9 r_4^{-3/2}\,\yr.
\ee

These estimates suggest that the three conditions given at the start of this
section are approximately satisfied in the radius interval $200\au\la r \la
6\times10^3\au$, a range of more than an order of magnitude. The actual range
of validity of our assumptions depends on the uncertain normalization and
radial dependence of the Oort-cloud mass distribution,

\subsection{Results}

\noindent
The remaining sections of this paper contain analytical results on the linear
stability of spherical near-Keplerian stellar systems, which we summarize
here. 

\begin{enumerate}

\item All such systems in which the \df\ is a decreasing function of angular 
momentum are stable to $l=1$ secular modes (\S\ref{sec:lopsided}).

\item All such systems in which the \df\ is an increasing function of angular
momentum that is non-zero at zero angular momentum are also stable to $l=1$
secular modes. 

\item Remarkably, all systems in which the \df\ is an increasing function of
angular momentum that is zero at zero angular momentum (i.e., the loss cone is
empty at all energies) are neutrally stable to an $l=1$ secular mode
(\S\ref{sec:neutral}). The spatial form of this mode corresponds to a uniform
displacement of the stellar system relative to the central mass. 

\item To explore the generality of our results, we have also examined flat,
non-rotating, near-Keplerian systems, which offer a foil to the spherical
systems. We find that flat systems do not support neutral modes, and that many
flat, non-rotating systems with empty loss cones are secularly unstable
(\S\ref{sec:flat}). This result suggests that flattening often transforms the
neutral mode in spherical systems to an unstable mode.

\item Numerical normal-mode calculations strongly suggest
that secular modes with $l>1$ are generally stable (\S\ref{sec:lbig}).

\end{enumerate}

\section{Linear perturbation theory}

\label{sec:linpert}

\subsection{The equilibrium system}

\noindent
We consider a spherical stellar system with equilibrium \df\ $F(\vv,\vx)$. By
``spherical'' we mean that the \df\ is invariant under spatial rotations of
the coordinate frame about the center of the system, which implies
that $F=F(r,v_r,v_t)$ where $r$ is the radius, $v_r>0$ is the radial speed,
and $v_t>0$ is the tangential speed. 

We introduce action-angle variables $(\vI,\vw)$, which are chosen so that
$I_2$ is the total angular momentum, $I_3$ is the $z$-component of the angular
momentum, and $I_1=J_r+I_2$ where $J_r$ is the radial action \citep{tw84}. The
allowed values for the actions (the action space) are given by
\be
0\le I_2 \le I_1,\quad -I_2\le I_3\le I_2.
\ee
The conjugate angle $w_3$ is the longitude of the node, that is, the azimuthal
angle at which the orbital plane intersects the equatorial plane; there are
two nodes separated by $\pi$, and $w_3$ is usually chosen to be the one at
which the orbit has $\dot z>0$ (the ascending node). The angle $w_1$ is the
orbital phase measured from pericenter, that is, $w_1=2\pi(t-t_p)/T_r$ where
$T_r$ is the radial period and $t_p$ is the time of pericenter passage. At
pericenter, $w_2$ is the angle measured in the orbital plane from the
ascending node. The interpretation of these actions and angles in a Keplerian
potential is described in \S\ref{sec:first}.

If the \df\ is written as a function of the action-angle variables,
then according to Jeans's theorem it can only depend on the actions
$\vI$ and on the longitude of the node $w_3$ since these are the only
integrals of motion in a spherical potential. Rotational invariance
requires in addition that the \df\ is independent of both $w_3$ and
the $z$-component of angular momentum $I_3$.  Thus $F=F(I_1,I_2)$.

\subsection{The linearized collisionless Boltzmann equation}

\noindent
We now subject the stellar system to a weak perturbing potential
$\Phi$. The resulting perturbation in the \df, $f$, is determined by
the linearized collisionless Boltzmann equation [eq.\ 
(\ref{eq:vlasov}), except now the phase space is defined by velocity
and position $(\vv,\vx)$ rather than momentum and position
$(\vp,\vx)$]. We restrict ourselves to the case in which the
perturbation is self-consistent, so that the perturbing potential
$\Phi$ is determined by the perturbed \df\ through Poisson's equation,
\be \Phi(\vx,t)=-G\int {d\vv'd\vx'\,f(\vv',\vx',t)\over|\vx-\vx'|}.
\label{eq:pois}
\ee
We assume for the moment that the mass $M$ is fixed at the origin; the 
consequences of lifting this assumption are addressed in \S\ref{sec:kep}.

We may assume that the time dependence of $f$ is proportional to
$\exp(\lambda t)$. Equation (\ref{eq:vlasov}) can then be written as a linear
eigenvalue equation,
\be
-\{f,H_0\} - \{F,\Phi[f]\}=\lambda f,
\label{eq:linop}
\ee
where $\Phi[f]$ is the linear operator on $f$ defined by equation
(\ref{eq:pois}). In action-angle variables this reads
\be
-{\p f\over\p\vw}\cdot{\p H_0\over\p\vI}+{\p
F\over\p\vI}\cdot{\p\Phi\over\p\vw}= -\vOmega\cdot{\p f\over\p\vw}+{\p
F\over\p\vI}\cdot{\p\Phi\over\p\vw}=\lambda f.
\ee

\subsection{Expansion in spherical harmonics}

\noindent
Consider a perturbing potential of the form
\be
\Phi_p(\vx)=\Phi_{lm}(r)Y_{lm}(\theta,\phi).
\label{eq:pot}
\ee
The mass density that produces this potential is
\be
\rho_p(\vx)=\rho_{lm}(r)Y_{lm}(\theta,\phi),
\ee
where 
\be
\Phi_{lm}(r)=-{4\pi G\over 2l+1}\int_0^\infty dr'\,(r')^2
\rho_{lm}(r'){r_<^l\over r_>^{l+1}},\quad \rho_{lm}(r)=\int \sin\theta d\theta
d\phi Y_{lm}^\ast(\theta,\phi)\rho_p(\vx), 
\label{eq:poisint}
\ee
and $r_<=\min(r,r')$, $r_>=\max(r,r')$. 

A spherical harmonic can be written in action-angle variables as \citep{tw84}
\be
Y_{lm}(\theta,\phi)=\sum_{j=-l}^li^{m-j}y_{lj}r^l_{jm}(\beta)
e^{i(j\chi+mw_3)}
\label{eq:sph}
\ee 
where \be y_{lj}\equiv Y_{lj}(\half\pi,0) \ee is zero unless $l-j$
is even, $r^l_{jm}(\beta)$ is a rotation matrix \citep{edm60},
$\beta=\cos^{-1}I_3/I_2$ is the inclination, and $\chi$ is the angle
measured in the orbital plane from the ascending node. Since $w_2$ is
a measure of the angle between the ascending node and pericenter, the
angle $\chi$ can be written as $w_2+\psi$ where $\psi$ is a function
of $I_1$, $I_2$, and the radial phase $w_1$ only (in \S\ref{sec:kep}
we shall identify $\psi$ with the true anomaly in Keplerian
potentials). Note that $y_{lj}$ and $r^l_{jm}(\beta)$ are both real
and $y_{lj}$ is non-zero only if $l-j$ is even. 
Since the radius $r$ is also a function of $I_1$, $I_2$, and $w_1$
only, and phase space is periodic in $w_1$, the potential
(\ref{eq:pot}) can be written in action-angle variables as 
\be
\Phi(\vx)=\sum_{l_1=-\infty}^\infty
\sum_{j=-l}^li^{m-j}y_{lj}r^l_{jm}(\beta)W^{l_1}_{ljm}(I_1,I_2)
e^{i(l_1w_1+jw_2+mw_3)},
\label{eq:pota}
\ee
where
\be
W^{l_1}_{ljm}={1\over 2\pi}\int_{-\pi}^\pi dw_1\,e^{i(j\psi-l_1w_1)}
\Phi_{lm}(r).
\label{eq:wdef}
\ee

An arbitrary \df\ can be expanded as a Fourier series in the angles,
\be
f(\vI,\vw)=\sum_\vl f_\vl(I_1,I_2,\beta)e^{i\vl\cdot\vw}.
\label{eq:ffour}
\ee
Using the expansions (\ref{eq:pota}) and (\ref{eq:ffour}), the
eigenvalue equation (\ref{eq:linop}) can be solved explicitly,
\be
f_\vl(\vI)=
{i^{1+l_3-l_2}y_{ll_2}r^l_{l_2l_3}(\beta)W^{l_1}_{ll_2l_3}(I_1,I_2)\over
\lambda +i\vl\cdot\vOmega(I_1,I_2)}\left(l_1{\p F\over\p I_1}+l_2{\p F\over\p
I_2}\right). 
\label{eq:exp}
\ee
This result shows that the perturbed \df\ arising from a potential
perturbation whose angular dependence is proportional to $Y_{lm}(\theta,\phi)$
must have the form
\be
f_{lm}(\vI,\vw)=\sum_{j=-l}^le^{i\phi_{ljm}}g_{ljm}(I_1,I_2,w_1)
r^l_{jm}(\beta)e^{i(jw_2+mw_3)};
\label{eq:sphxxx}
\ee
here the factor 
\be
e^{i\phi_{ljm}}\equiv i^{m-j}{|y_{lj}|\over y_{lj}}
\ee
is added to simplify later formulae.  

We may now show that the density or potential arising from the response
$f_{lm}$ must have angular dependence proportional to
$Y_{lm}(\theta,\phi)$. The response potential having angular dependence
proportional to $Y_{nk}(\theta,\phi)$ is given by equation (\ref{eq:poisint})
as
\be
\Phi_{nk}(r)=-{4\pi G\over 2n+1}\int
d\vv'd\vx'\,Y_{nk}^\ast(\theta',\phi')f_{lm}(\vv',\vx'){r_<^n\over r_>^{n+1}}.
\label{eq:poisinta}
\ee
Since phase-space volume is invariant under canonical transformations, we may
replace $d\vv' d\vx'$ by $d\vI'd\vw'$ and substitute equations (\ref{eq:sph})
and (\ref{eq:sphxxx}) for $Y_{nk}$ and $f_{lm}$. Noting that the rotation
matrices are real, and using the orthogonality relation
\be
\int_0^\pi\sin\beta d\beta r^l_{jm}(\beta)r^{l'}_{jm}(\beta)={2\over
2l+1}\delta_{ll'},
\label{eq:orthog}
\ee
we obtain
\be
\Phi_{nk}(r)=-{2^5\pi^3 G\over
(2l+1)^2}\delta_{nl}\delta_{km}\sum_{j=-l}^l|y_{lj}|\int 
dI_1'I_2'dI_2'dw_1'\, 
e^{-ij\psi'}g_{ljm}(I_1',I_2',w_1'){r_<^l\over r_>^{l+1}}.
\label{eq:poisintc}
\ee
Thus the response potential (or density) is zero unless $n=l$ and $k=m$.  In
other words a perturbing potential $\propto Y_{lm}$ produces a response
potential with the same angular dependence, a consequence of the rotational
invariance of the unperturbed \df\ and potential.

This result also shows that the eigenfunctions of Poisson's equation
and the linearized collisionless Boltzmann equation (eqs.\ 
\ref{eq:pois} and \ref{eq:linop}) can be chosen to have specific
angular dependences: the density and potential can be chosen equal to
functions of radius times $Y_{lm}(\theta,\phi)$, and the \df\ can be
chosen to be a sum of terms that are functions of $I_1$, $I_2$ and
$w_1$ times $r^l_{jm}(\beta)\exp[i(jw_2+mw_3)]$ with $|j|\le l$.
Moreover, the eigenvalues $\lambda$ must be independent of the
orientation of the coordinate system and hence must be independent of
$m$.

\section{Near-Keplerian stellar systems}

\label{sec:kep}

\noindent
We consider motion in an unperturbed Hamiltonian $H_0(\vv,\vx)=
\half v^2-GM/r+\Phi_\ast(r)$, where $M$ is the mass of the
central object and $\Phi_\ast(r)$ is the potential due to a spherical stellar
system centered on $M$. The mass of the stellar system interior to $r$ is
$M_\ast(r)=4\pi\int_0^r\,dr\,r^2\rho_\ast(r)$, and we assume that
$M_\ast(r)/M\sim\epsilon\ll1$, so that the potential is nearly
Keplerian. Since the equilibrium system is spherical,
$GM_\ast(r)/r^2=d\Phi_\ast/dr$.

\subsection{Near-Keplerian motion}

\label{sec:first}

Since the potential is nearly Keplerian, the actions can be written
approximately in terms of the standard Keplerian orbital elements
\be
I_1\simeq (GMa)^{1/2},\qquad I_2\simeq 
(GMa)^{1/2}(1-e^2)^{1/2}, \qquad I_3=I_2\cos \beta,
\label{eq:acdef}
\ee
where $a$ is the semi-major axis, $e$ is the eccentricity, and $\beta$ is the
inclination.  The conjugate angles are
\be
w_1\simeq \ell,\qquad  w_2\simeq\omega, \qquad  w_3\simeq\Omega,
\label{eq:angledef}
\ee
where $\ell$ is the mean anomaly, $\omega$ is the argument of pericenter, and
$\Omega$ is the longitude of the ascending node. If $\chi$ is the
angle measured in the orbital plane from the ascending node
(eq.~\ref{eq:sph}), then $\psi=\chi-\omega$ is the true anomaly. 

The Hamiltonian can be written
\be
H_0=-{(GM)^2\over2I_1^2}+\Phi_\ast.
\ee
The unperturbed frequencies
$\vOmega\equiv\dot\vw= \p H_0/\p \vI$ are
\be
\Omega_1\simeq \left(GM/a^3\right)^{1/2}, \qquad \Omega_2\simeq
0, \qquad \Omega_3=0.
\ee

The frequency $\Omega_2$ is equal to the time-averaged value of the precession
rate $\dot\omega$, and is smaller than $\Omega_1$ by a factor of order
$\epsilon\sim M_\ast(r)/M\ll1$. Although small, $\Omega_2$ plays a central
role in determining the secular stability of near-Keplerian systems. The
simplest way to determine $\Omega_2$ is through Gauss's method, which states
that in a spherical, near-Keplerian potential
\be
\dot\omega={(1-e^2)^{1/2}\over\Omega_1ae}{d\Phi_\ast\over dr}\cos \psi.
\ee
Thus
\be
\Omega_2=\langle\dot\omega\rangle=
{(1-e^2)^{1/2}\over\Omega_1ae}\left\langle{d\Phi_\ast\over dr}\cos
\psi \right\rangle=
{G(1-e^2)^{1/2}\over\Omega_1ae}\left\langle{M_\ast(r)\cos
\psi\over r^2}\right\rangle,
\label{eq:pomegadot}
\ee
where $\langle X\rangle=(2\pi)^{-1}\int_0^{2\pi}dw_1\,X$ represents a time
average over the orbit. We can rewrite the time average as an average over
true anomaly: since $dw_1=\Omega_1dt$, and the angular momentum 
$I_2=r^2d\psi/dt$, 
\be
\langle X\rangle={\Omega_1\over 2\pi
I_2}\int_0^{2\pi}d\psi\, r^2X={1\over2\pi a^2(1-e^2)^{1/2}}\int_0^{2\pi}
d\psi\,r^2X. 
\label{eq:timeav}
\ee
Using this result in equation (\ref{eq:pomegadot}) we have
\be
\Omega_2={\Omega_1\over \pi Me}\int_0^{\pi}d\psi\, M_\ast(r)\cos \psi.
\ee

We now investigate the sign of $\Omega_2$. We may write
\be
\int_0^{\pi} d\psi\,M_\ast(r)\cos \psi=\int_0^{\pi/2}d\psi\,\cos \psi
\{M_\ast[r(\psi)]-M_\ast[r(\pi-\psi)]\}.
\label{eq:ineq}
\ee
For $0\le \psi<\half\pi$, $r(\pi-\psi)$ is greater
than $r(\psi)$. Since $M_\ast(r)$ is an increasing
function of $r$, the integrand in equation (\ref{eq:ineq}) is negative so
$\Omega_2$ is also negative. Thus the line of apsides precesses in
the opposite direction to the orbital motion, so long as the precession is
dominated by the self-gravity of a spherical system.

\subsection{Secular oscillations}

We now examine secular oscillations in near-Keplerian systems; as
described in \S\ref{sec:times}, these have characteristic growth rates
or frequencies of order $\Omega_2 \sim\epsilon\Omega_1\ll\Omega_1$.

We shall work in a frame centered on the mass $M$. Then in addition to the
perturbing potential $\Phi$ described by Poisson's equation (\ref{eq:pois})
there is also an indirect potential $\Phi^i$ arising from the acceleration of
the frame, which is given by
\be
\Phi^i(\vx,t)={G\vx\over M}\cdot\int \vx'{d\vx'\over|\vx'|^3}\rho(\vx',t),
\ee
where $\rho=\int d\vv\, f$. However, we now show that the indirect potential
can be neglected. Since we are interested in secular perturbations, the
perturbed density $\rho$ arises from the superposition of the time-averaged
density along individual orbits. For any Kepler orbit
\be
\left\langle {\vx'\over|\vx'|^3}\right\rangle=
\left\langle{\cos \psi\over r^2}\right\rangle \vp,
\ee
where $\psi$ is the true anomaly and $\vp$ is a unit vector pointing toward
pericenter. This time average is easily seen to vanish, from the last
expression in equation (\ref{eq:timeav}).  Thus the indirect potential is
zero for secular oscillations, in the approximation at which we are
working. Terms that are of higher order in the small parameter
$\epsilon$ are responsible for the weak acceleration of the central
mass that is required so that the center of mass of $M$ and $M_\ast$ 
remains fixed.

Equation (\ref{eq:exp}) provides a formal solution to the linearized
collisionless Boltzmann equation. To solve Poisson's equation, the
perturbed \df\ $f_\vl$ must be of order the perturbed potential
$W^{l_1}_{ll_2l_3}$. However, the unperturbed \df\ $F(I_1,I_2)$ is of
order $\epsilon$, as is the eigenvalue $\lambda$ and the precession
rate $\Omega_2$, while the frequency $\Omega_1$ is independent of
$\epsilon$. Therefore, self-consistency requires that only terms with
$l_1=0$ are important; in other words, the perturbed \df\ may be taken
to be independent of $w_1$ and only the orbit-averaged component of
the perturbing potential $W^0_{ll_2l_3}$ is important. The same
approximations underlie secular perturbation theory in celestial
mechanics.

Using the expressions (\ref{eq:pota}) and (\ref{eq:sphxxx}) for the perturbed
potential and \df, and restricting the eigenvalue equation (\ref{eq:linop}) to
terms with $l_1=0$ yields
\be
-ij\Omega_2g_{ljm}+ij|y_{lj}|W^0_{ljm}{\p F\over\p I_2}=\lambda g_{ljm},
\label{eq:ggg}
\ee
where $g_{ljm}$ is a function of $I_1$ and $I_2$; from equation (\ref{eq:wdef})
\be
W^{0}_{ljm}={1\over 2\pi}\int_{-\pi}^\pi dw_1e^{ij\psi}
\Phi_{lm}(r)=\langle\Phi_{lm}(r)\cos j\psi\rangle;
\label{eq:wdefa}
\ee
and $\Phi_{lm}$ is related to $g_{ljm}$ by equation (\ref{eq:poisintc}):
\be
\Phi_{lm}(r)=-{2^5\pi^3 G\over(2l+1)^2}\sum_{j'=-l}^l|y_{lj'}|\int 
dI_1'I_2'dI_2'\,
g_{lj'm}(I_1',I_2')\int dw_1'e^{-ij'\psi'}{r_<^l\over r_>^{l+1}}.
\label{eq:poisintd}
\ee
These equations can be rewritten as
\be
-ij\Omega_2g_{lj}+ij{\p F\over\p I_2}\sum_{j'=-l}^l 
T^{l}_{jj'}(g_{lj'})=\lambda g_{lj};
\label{eq:ggga}
\ee
here we have dropped the subscript $m$ on $g_{ljm}$ since the eigenvalue
equation is independent of $m$. We have also defined the operator
\be
T^{l}_{jj'}(z)=-{2^4\pi^2G\over (2l+1)^2}|y_{lj}||y_{lj'}|\int 
dI_1'I_2'dI_2'\,z(I_1',I_2')\int dw_1dw_1'e^{i(j\psi-j'\psi')}
{r_<^l\over r_>^{l+1}};
\label{eq:poisinte}
\ee
here $r$ and $\psi$ are functions of $I_1$, $I_2$, and $w_1$; $r'$ and
$\psi'$ are functions of $I_1'$, $I_2'$, and $w_1'$; and as usual
$r_>=\max(r,r')$ and $r_<=\min(r,r')$.  Note that equation (\ref{eq:ggga})
implies that  $g_{l0}=0$.

The eigenvalue equation (\ref{eq:ggga}) is linear in the eigenfrequency
$\lambda$. This is in contrast to the usual equation for normal
modes of stellar systems, which is nonlinear in the frequency (e.g.,
\citealt{kal77,wein91}). Eigenvalue equations that are linear in the frequency
have simpler analytic properties and are easier to solve numerically. For disk
systems, \cite{pol04} has stressed that such equations arise whenever the
response is dominated by a single resonance, that is, by one integer triple
$\vl$ in the Fourier expansion in the actions. The result (\ref{eq:ggga}) is
slightly more general, since several resonances with different values of
$l_2=j$ (but the same values of $l_1=0$ and $l_3=m$) are involved.

\subsection{The Hermitian form of the eigenvalue equation}

\label{sec:hermit}

\noindent
Equations (\ref{eq:ggga}) and (\ref{eq:poisinte}) completely describe the
linearized secular oscillations of spherical near-Keplerian stellar
systems. We now show that this eigenvalue equation can be rewritten in a
Hermitian form, so long as $\p F/\p I_2$ has the same sign everywhere.

Equation (\ref{eq:poisintd}) determines the potential $\Phi_{lm}(r)$ in terms
of the \df\ specified by $g_{lj}$. The true anomaly $\psi$ and the radius $r$
are both even functions of $w_1$. Moreover $|y_{lj}|$ is even in $j$. Thus the
potential depends on the \df\ only through the combination $g_{lj}+
g_{l,-j}$. Therefore it is natural to rewrite (\ref{eq:ggga}) in terms of the
combinations\footnote{This approach is closely related to Antonov's (1960)
trick of analyzing stability of spherical systems by forming
\dfs\ that are even and odd in the velocities.}
\be
g^+_{lj}(I_1,I_2)=\half[g_{lj}(I_1,I_2)+g_{l,-j}(I_1,I_2)]\quad , \quad
g^-_{lj}(I_1,I_2)=\half[g_{lj}(I_1,I_2)-g_{l,-j}(I_1,I_2)].
\ee
Since the operator $T^l_{jj'}$ is an even function of $j$ and $j'$, we find
\be
-ij\Omega_2g^+_{lj}+ij{\p F\over\p I_2}\sum_{j'=-l}^l T^{l}_{jj'}(g^+_{lj'})
=\lambda g^-_{lj},
\qquad
-ij\Omega_2g^-_{lj}=\lambda g^+_{lj}.
\label{eq:gggp}
\ee
Since the second equation implies that $g^+_{l0}=0$, the sum in the first
equation can be restricted to $j'\not=0$. The two equations can be combined
into a single eigenvalue equation, 
\be
\lambda^2g^+_{lj}=-j^2\Omega_2^2g^+_{lj}+ j^2\Omega_2{\p F\over\p I_2}
\sum_{j'=-l\atop j'\not=0}^l T^{l}_{jj'}(g^+_{lj'})\equiv S^l_j(g^+_{lj'}).
\label{eq:eigenb}
\ee

We now restrict ourselves to stellar systems in which $\p F/\p I_2$ is
non-zero and has the same sign everywhere. Since $\Omega_2<0$
(\S\ref{sec:first}) we may define an inner product
\be
[\va,\vb]\equiv -\sum_{j=-l\atop j\not=0}^l\int 
{dI_1I_2dI_2\over j^2\Omega_2|\p F/\p I_2|} 
a^\ast_jb_j;
\label{eq:inner}
\ee
then $\vS^l$ is a Hermitian operator, since
\label{eq:hermit}
\begin{eqnarray}
&[\va,\vS^l(\vb)]&=\sum_{j=-l\atop j\not=0}^l
\int {dI_1I_2dI_2\Omega_2\over|\p F/\p I_2|} a^\ast_jb_j \nonumber\\
&&\hspace{-0.5cm}
+{2^4\pi^2G\epsilon\over (2l+1)^2}\sum_{j,j'=-l\atop j,j'\not=0}^l
|y_{lj}||y_{lj'}|\int dI_1I_2dI_2\,
a^\ast_j\int dI_1'I_2'dI_2'\,b'_{j'}
\int dw_1dw_1'\,e^{i(j\psi-j'\psi')}{r_<^l\over r_>^{l+1}}\nonumber \\
&&=[\vb,\vS^l(\va)]^\ast,
\label{eq:sdef}
\end{eqnarray}
where $\epsilon=\mbox{sgn}(\p F/\p I_2)$. Note that the domain of $\vS^l$ is
restricted to vectors $\va$ with $a_{-j}=a_j$.

The second term in equation \ref{eq:hermit} is directly proportional to the
mutual gravitational potential energy of the phase-space densities defined
by $\va$ and $\vb$.  

Since $\vS^l$ is Hermitian, its eigenvalues are real and its eigenfunctions
can be chosen to be orthogonal. Thus $\lambda^2$ is real, and all modes have
one of two forms: either pairs of damped and growing modes ($\lambda^2>0$) or
pairs of oscillatory modes ($\lambda^2<0$). Some or all of the oscillatory
modes may be singular or van Kampen modes, as we discuss further in
\S\ref{sec:oscill}. 

The stability properties of $\vS^l$ can be investigated analytically for
$l=1$, as shown in the following subsections. A numerical investigation of
stability for $l>1$ is described in \S\ref{sec:lbig}. 

\subsection{Lopsided $(l=1)$ modes}

\label{sec:lopsided}

\noindent
The simplest disturbances are those with $l=1$; as the arguments in
\S\ref{sec:analog} suggest, these are also the most likely to
be unstable. For $l=1$ the operator $\vS^l$ can be simplified. 

It is useful to consider the inner product $[\va,\vS^1(\va)]$. We have
\be
y_{1\pm1}=\mp\left(3\over 8\pi\right)^{1/2}\quad,\quad y_{10}=0,
\label{eq:ylmone}
\ee
and $a_{-1}=a_1$. Thus
\begin{eqnarray}
[\va,\vS^1(\va)]&=&2\int {dI_1I_2dI_2\,\Omega_2\over |\p F/\p
I_2|} |a_1|^2 \nonumber\\
&&+{8\pi G\epsilon\over 3}\int 
dI_1I_2dI_2\,a^\ast_1\int dI_1'I_2'dI_2'
a'_1\int dw_1dw_1'\,\cos\psi\cos\psi'{r_<\over r_>^2}\nonumber \\
&\equiv& S_\alpha+S_\beta.
\label{eq:sdefa}
\end{eqnarray}

According to equation (\ref{eq:pomegadot}), 
\be
\Omega_2={I_2\over 2\pi Me}\int dw_1 {M_\ast(r)\cos\psi\over r^2}=
{I_2\Omega_1\over \pi Me}\int dr{M_\ast(r)\cos\psi\over v_rr^2};
\ee
in the last expression we have replaced the integral over the angle $w_1$ from
0 to $2\pi$ with twice the integral over the radius $r$ from pericenter to
apocenter, using the relation $dw_1=\Omega_1dr/v_r$. The radial speed $v_r$
is a positive function of $I_1$, $I_2$ and $r$ defined by
$v_r=[2H_0(I_1,I_2)+2GM/r-I_2^2/r^2]^{1/2}$. The first term in equation
(\ref{eq:sdefa}) then becomes
\be
S_\alpha={2\over\pi M}\int dr {M_\ast(r)\over r^2}\int
{dI_1I_2^2dI_2\,\Omega_1\over ev_r|\p F/\p I_2|} \cos\psi |a_1|^2.
\ee
Using the relation
\be
{\p v_r\over\p r}={GM\over r^2v_r}\left({I_2^2\over
GMr}-1\right)={GMe\cos\psi\over r^2v_r},
\label{eq:partial}
\ee
this expression can be rewritten as 
\be
S_\alpha={2\over\pi GM^2}\int drM_\ast(r){\p\over\p r}\int
{dI_1I_2^2dI_2\,\Omega_1\over e^2|\p F/\p I_2|}v_r |a_1|^2,
\ee
where the inner integral is over all values of $I_1$ and $I_2$ for which $v_r$
is real at a given radius. We may now integrate by parts; the boundary
terms vanish since $M_\ast=0$ at $r=0$, and since we may assume that the
perturbation described by $a_1$ vanishes at sufficiently large distances. Thus
\be
S_\alpha=-{8\over GM^2}\int dr\, r^2\rho_\ast(r)\int
{dI_1I_2^2dI_2\,\Omega_1\over e^2|\p F/\p I_2|}v_r |a_1|^2,
\label{eq:sadef}
\ee
where $\rho_\ast(r)$ is the density of the equilibrium stellar system.

Similar manipulations on the second term of equation (\ref{eq:sdefa}) yield
\be
S_\beta={32\pi\epsilon\over 3GM^2}\int dr dr'\, w(r,r'){\p^2\over\p r\p r'}
\int {dI_1I_2dI_2\,\Omega_1\over e}v_ra^\ast_1
\int {dI_1'I_2'dI_2'\,\Omega_1'\over e'}v_r'a'_1,
\ee
where $w(r,r')=(rr')^2r_</r_>^2=r_<^3$. 
Integrating by parts with respect to $r$ and $r'$, assuming that the
boundary terms vanish for the reasons given in the preceding paragraph, and
using the result $\p^2w(r,r')/\p r\p r'=3r^2\delta(r'-r)$, we find
\be
S_\beta={32\pi\epsilon \over GM^2}\int dr\,r^2 \left|\int
{dI_1I_2dI_2\,\Omega_1\over e}v_ra_1\right|^2. 
\label{eq:sbdef}
\ee

First consider systems in which the equilibrium \df\ is a decreasing function
of angular momentum, $\p F/\p I_2<0$. Then $\epsilon=\mbox{sgn}(\p F/\p
I_2)=-1$ and both $S_\alpha$ (eq.\ [\ref{eq:sadef}]) and $S_\beta$ (eq.\
[\ref{eq:sbdef}]) are negative. Thus $[\va,\vS^1(\va)]<0$, so the operator
$\vS^1$ is negative, and all of its eigenvalues are less than zero. Thus all
spherical near-Keplerian stellar systems with $\p F/\p I_2<0$ are stable to
$l=1$ secular perturbations. 

Determining stability when $\p F/\p I_2>0$ requires more work. We use
Schwarz's inequality,
\be
\int dI_1dI_2\,|A|^2\int dI_1dI_2\,|B|^2\ge \left|\int dI_1dI_2\,AB\right|^2,
\ee
with 
\be
A={I_2a_1\over e}\left(\Omega_1 v_r\over\p F/\p I_2\right)^{1/2},\qquad 
B=\left(\Omega_1 v_r\p F/\p I_2\right)^{1/2}.
\label{eq:defab}
\ee
Then
\be
\int dI_1dI_2\,|B|^2=\int dI_1dI_2\,\Omega_1v_r{\p F\over\p I_2}=
-\int dI_1\Omega_1Fv_r|_{I_2=0}+{1\over r^2}\int{dI_1I_2dI_2\,\Omega_1\over
v_r}F;
\ee
in the last expression we have integrated by parts and used the facts
that $\Omega_1$ is independent of $I_2$ and 
\be
{\p v_r\over \p I_2}=-{I_2\over r^2v_r}.
\label{eq:ident}
\ee
The final integral in this expression is simply related to the stellar
density $\rho_\ast(r)$. To show this, we change integration variables from
$(I_1,I_2)$ to $(v_r,v_t)$ where $v_t$ is the tangential speed. We have
$I_2=rv_t$, and at constant $I_2$ and $r$ the differential energy is
$dE=\Omega_1dI_1=v_rdv_r$ so
\be
dI_1={v_rdv_r\over\Omega_1}.
\label{eq:didvr}
\ee
Thus
\be
{1\over r^2}\int{dI_1I_2dI_2\,\Omega_1\over v_r}F=\int dv_rv_tdv_tF={1\over
4\pi}\rho_\ast,
\label{eq:xxxyyy}
\ee  
and
\be
\int dI_1dI_2|B|^2\le {1\over4\pi}\rho_\ast,
\ee
with equality if and only if the \df\ vanishes at zero angular
momentum. Schwarz's inequality then becomes
\be
\int{dI_1I_2^2dI_2\,\Omega_1\over e^2\p F/\p I_2}v_r |a_1|^2\ge
{4\pi\over\rho_\ast}\left|\int {dI_1I_2dI_2\,\Omega_1\over e}v_ra_1\right|^2.
\ee
Substituting this inequality into equations (\ref{eq:sadef}) and
(\ref{eq:sbdef}) yields
\be
[\va,\vS^1(\va)]=S_\alpha+S_\beta\le 0.
\ee
Thus the operator $\vS^1$ is non-positive, and all of its eigenvalues are less
than or equal to zero. We conclude that all spherical near-Keplerian stellar
systems with $\p F/\p I_2>0$ are stable or neutrally stable to secular $l=1$
perturbations.

\subsection{The neutral $l=1$ mode}

\label{sec:neutral}

\noindent
The maximum of $[\va,\vS^1(\va)]$ is zero, which is achieved when
$F(I_1,I_2=0)=0$, $\p F/\p I_2>0$, and $A=B$. This bound is achieved if and
only if $\va$ is the eigenvector with the largest eigenvalue,
$\lambda^2=0$. Thus all near-Keplerian stellar systems with a \df\ that is an
increasing function of angular momentum and vanishes at zero angular momentum
support a neutral $l=1$ mode with
\be
g^+_{11}\propto {e\over I_2}{\p F\over\p I_2}.
\ee
To convert this to a \df, we shall assume that $m=0$, that is, we orient the
coordinate axes so that the $l=1$ mode is axisymmetric. The rotation
matrices are 
\be
r^1_{\pm 1,0}(\beta)=\mp{1\over  2^{1/2}}\sin\beta,
\label{eq:rotone}
\ee
and equation (\ref{eq:sphxxx}) yields
\be
f_{10} \propto {e\over I_2}{\p F\over\p I_2}\sin\beta\sin\omega.
\label{eq:dfdef}
\ee

The density corresponding to this \df\ is straightforward to determine. The
eccentricity vector $\ve$, which points toward pericenter and has magnitude
equal to the eccentricity, is defined by
\be
\ve={1\over GM}\vv\times(\vrr\times\vv)-\hat{\vrr}.
\label{eq:evec}
\ee
The $z$-component of the eccentricity vector is given in terms of the
orbital elements by 
\be
\ve\cdot\hat\vz=e\sin\beta\sin\omega
\label{eq:eveca}
\ee
and in terms of the velocity by
\be
\ve\cdot\hat\vz={r\over GM}(v^2\cos\theta-s_rs_zv_rv_z)-\cos\theta,
\ee
where $\theta$ is the usual polar angle, $v$ is the total speed, $v_r>0$ and
$v_z>0$ are its radial and $z$-components, and $s_r$ (and $s_z$) are $\pm 1$
to account for in-and-out (and up-and-down) motion. Thus the \df\
(\ref{eq:dfdef}) yields the density
\be
\rho_{10}=\int d\vv f_{10}\propto \int {d\vv \over I_2}{\p F\over\p I_2}
\left[{r\over GM}(v^2\cos\theta-s_rs_zv_rv_z)-\cos\theta\right].
\ee
The equilibrium \df\ $F(I_1,I_2)$ depends on the velocity only through $v_r$
and $v_t$, where $v_t$ is the tangential speed. If we define $\chi$ to be
the angle between the tangential component of the velocity vector and the unit
vector $\vphihat$ in spherical polar coordinates, then
$s_zv_z=s_rv_r\cos\theta+v_t\sin\chi\sin\theta$ and we have
\be
\rho_{10}\propto \int {dv_r v_t dv_t d\chi\over I_2}{\p F\over\p I_2}
\left[{r\over
GM}(v_t^2\cos\theta-s_rv_rv_t\sin\chi\sin\theta)-\cos\theta\right]. 
\ee
The term proportional to $\sin\chi$ integrates to zero; thus,
\be
\rho_{10}\propto \cos\theta \int {dv_r v_t dv_t \over I_2}{\p F\over\p I_2}
\left({rv_t^2\over GM}-1\right).
\ee 
Using equation (\ref{eq:didvr}) and the relation $rv_t=I_2$, this becomes
\be
\rho_{10}\propto {\cos\theta\over r^2} \int {dI_1dI_2\Omega_1\over v_r}{\p
F\over\p I_2} \left({I_2^2\over GMr}-1\right).
\ee 
Using equation (\ref{eq:partial}) this simplifies to
\be
\rho_{10}\propto \cos\theta {\p\over \p r}\int {dI_1dI_2\Omega_1}{\p
F\over\p I_2}v_r.
\ee 
We now integrate by parts with respect to $I_2$ and assume that
$F(I_1,I_2=0)=0$ since this is necessary for the existence of a neutral mode;
then using the identity (\ref{eq:ident})
\be
\rho_{10}\propto \cos\theta {\p\over \p r}{1\over r^2}\int 
{dI_1I_2dI_2\Omega_1\over v_r} F\propto \cos\theta {d\rho_\ast\over dr},
\label{eq:last}
\ee  
where the last expression follows from equation (\ref{eq:xxxyyy}). Thus the
density in the neutral mode is obtained by a uniform displacement of the
equilibrium stellar density. 

The neutral mode that we have found is reminiscent of the trivial mode that is
present in isolated stellar systems when the origin of the coordinate system
is displaced from the center of the equilibrium system; in spherical systems
this displacement mode is an $l=1$ neutral mode with density $\propto
\cos\theta d\rho_\ast/dr$, just as in equation (\ref{eq:last}). However, the
mode derived here is quite different, for several reasons. First, it
represents a displacement of the center of the stellar system from the point
mass that dominates the potential rather than from an arbitrary coordinate
origin (recall that we are working in the frame centered on $M$). Second,
although the spatial structure corresponds to a uniform displacement, the
phase-space structure does not.  Third, the neutral displacement mode is
present in all isolated equilibrium stellar systems, whereas the mode we have
found here is present only in spherical systems: for example, disk systems do
not support such a mode (\S\ref{sec:flat}).

\section{Discussion}

\subsection{A short derivation of the neutral mode}

\noindent
There is a simpler proof of the presence of a neutral mode in a wide range of
near-Keplerian spherical systems.

We may orient the coordinate system so that any $l=1$ potential perturbation
is axisymmetric, $\Phi_p(\vx)=\Phi_{10}(r)Y_{10}(\theta,\phi)$. According to
equation (\ref{eq:pota}) the secular component of this potential can be
written as
\be
\Phi_s(\vx)=\sum_{j=\pm1}i^{-j}y_{1j}r^1_{j0}(\beta)W^0_{1j0}(I_1,I_2)
e^{ijw_2}.
\ee
Using equations (\ref{eq:wdef}), (\ref{eq:ylmone}), and (\ref{eq:rotone}),
this can be simplified to
\be
\Phi(\vx)=\left(3\over 4\pi\right)^{1/2}\langle\Phi_{10}(r)\cos\psi\rangle
\sin\beta\sin \omega.
\label{eq:phisec}
\ee
Thus the secular Hamiltonian is
\be
H_s(I_1,I_2,\omega)=H_0(I_1,I_2)+\left(3\over
4\pi\right)^{1/2}\langle\Phi_{10}(r)\cos\psi\rangle \sin\beta\sin \omega.
\label{eq:hamsec}
\ee

In a neutral mode, both the Hamiltonian and the \df\ are
time-independent. Since the secular Hamiltonian is time-independent it is an
integral of the motion; the other integrals are $I_1$ and $I_3$ (since the
secular Hamiltonian is independent of $w_1$ and $w_3$). Since the
\df\ is time-independent, Jeans's theorem implies
that it can only depend on the integrals of motion, so
$f=f(H_s,I_1,I_3)$. Since the perturbing potential is small we may write
\be
f\simeq f(H_0,I_1,I_3)+{\p f\over\p H_0}\left(3\over
4\pi\right)^{1/2}\langle\Phi_{10}(r)\cos\psi\rangle \sin\beta\sin \omega.
\ee
The unperturbed \df\ is $F(I_1,I_2)$ and this must equal
$f(H_0,I_1,I_3)$. Since $H_0$ depends only on $I_1$ and $I_2$, the explicit
dependence of $f$ on $I_3$ can be dropped and we must have
\be
{\p F\over\p I_2}={\p f\over\p H_0}{\p H_0\over\p I_2}=\Omega_2{\p f\over\p
H_0}. 
\ee
Thus the perturbed \df\ is 
\be
f_{10}\equiv f(H_s,I_1)-f(H_0,I_1)=
\left(3\over 4\pi\right)^{1/2}{\p F\over\p I_2}
{\langle\Phi_{10}(r)\cos\psi\rangle\over\Omega_2}\sin\beta\sin \omega.
\label{eq:dfpert}
\ee

Now let us assume that the perturbed potential arises from a uniform
displacement of the unperturbed stellar potential $\Phi_\ast(\vx)$ by an amount
$\xi$ in the $z$-direction. Then the perturbed potential is 
\be
\Phi_p(\vx)=-\xi\cos\theta{d\Phi_\ast\over dr}=\Phi_{10}(r)Y_{10}(\theta,\phi)
 \qquad\hbox{where}\qquad
\Phi_{10}(r)=-\left(4\pi\over 3\right)^{1/2}\xi{d\Phi_\ast\over dr}.
\label{eq:phidisp}
\ee
Combining this result with equation (\ref{eq:pomegadot}) for $\Omega_2$,
equation (\ref{eq:dfpert}) simplifies to
\be
f_{10}=-\xi {GM\over I_2}{\p F\over\p I_2}e\sin\beta\sin\omega.
\ee
Following the derivation in equations (\ref{eq:evec})--(\ref{eq:last}), the
corresponding density is
\be
\rho_p(\vx)=\int d\vv f = -\xi\cos\theta{d\rho_\ast\over dr}.
\ee
where as usual we have assumed that the equilibrium \df\ vanishes at zero
angular momentum. Comparing this result to equation (\ref{eq:phidisp})
shows that the response density gives rise to the perturbing potential, so we
have found a self-consistent normal mode. 

\subsection{Oscillatory modes}

\label{sec:oscill}

\noindent
In this subsection we discuss the oscillatory secular modes of spherical
near-Keplerian stellar systems, which have $\omega^2\equiv-\lambda^2>0$. Once
again, for simplicity we restrict ourselves to lopsided modes ($l=1$). Since
the operator $\vS^1$ is Hermitian, the set of eigenvalues $\omega^2$ (the
spectrum of $-\vS^1$) lies on the real line, and since $-\vS^1$ is
non-negative its spectrum lies in the interval $[0,\infty)$.

The properties of the normal modes of stellar systems have been investigated
by \citet{mat90}. In general there is a continuous spectrum, one or more
segments of the real line that are filled with the eigenvalues of singular
normal modes analogous to the van Kampen modes of plasma physics. The singular
modes that are excited by physical perturbations decay by Landau
damping. There may also be isolated eigenvalues in the gaps of the real line
that do not belong to the continuous spectrum, and these correspond to
oscillating modes that do not decay.

According to Mathur, the continuous spectrum is the set of all frequencies
$\omega^2$ spanned by the function $(\vl\cdot\vOmega)^2$ over the domain of
actions for which $F(\vI)$ is non-zero. For secular $l=1$ modes in spherical
systems, this function simplifies to $\Omega_2^2$. Thus if $\Omega_2^2$ ranges
continuously from $\Omega_{\rm min}^2$ to $\Omega_{\rm max}^2$ but not outside
this range, isolated oscillatory modes must either have
$0<\omega^2<\Omega_{\rm min}^2$ or $\omega^2>\Omega_{\rm max}^2$.

There are further constraints on the frequencies of isolated modes. A mode
with frequency $\omega^2$ satisfies $[\va,\vS^1(\va)]=-\omega^2[\va,\va]$. 
Using equations (\ref{eq:inner}) and (\ref{eq:sdefa}),
\be
S_\beta=2\int {dI_1I_2dI_2\over\Omega_2|\p F/\p
I_2|}(\omega^2-\Omega_2^2)|a_1|^2. 
\ee
If $\omega^2>\Omega_{\rm max}^2$ then the right side must be negative (recall
that $\Omega_2<0$). However, equation (\ref{eq:sbdef}) shows that $S_\beta$
has the same sign as $\epsilon$, which is the sign of $\p F/\p I_2$. Thus
isolated oscillatory modes in systems with $\p F/\p I_2>0$ must have
$0<\omega^2<\Omega_{\rm min}^2$, while if $\p F/\p I_2<0$ the modes must have
$\omega^2>\Omega_{\rm max}^2$. 

Let us examine further the systems with $\p F/\p I_2>0$. Since $|\Omega_2|\to
0$ as $I_2\to 0$ (eq.\ [\ref{eq:pomegadot}]), isolated oscillatory modes can
only exist if the stellar system has a minimum angular momentum or maximum
eccentricity at every energy, i.e., if the loss cone is empty. It is
straightforward to show that if the stellar system has density
$\rho_\ast\propto r^{-\gamma}$ then the precession frequency $\Omega_2\propto
a^{3/2-\gamma}$ at fixed eccentricity. Thus stellar systems with a minimum
semi-major axis (set, for example, by collisional destruction of stars), an
empty loss cone, and $\gamma\le 1.5$ can have a gap in the continuous spectrum
between $\Omega_{\rm min}$ and zero, in which isolated oscillatory modes may
occur.  We have searched numerically for such modes in near-Keplerian stellar
systems with the \df\
\be
f(I_1,I_2)\propto \cases{I_1^{b}\log(I_2/hI_1) &if $I_{\rm min}\le
I_1\le I_{\rm max}$ and $I_2>hI_1$;\cr
            0              &otherwise.\cr}
\label{eq:dfdefa}
\ee
this represents a stellar system in which the loss cone is empty whenever the
angular momentum is less than a fraction $h$ of the angular momentum of a
circular orbit of the same energy. If $I_{\rm max}\gg I_{\rm min}$, the
corresponding density is roughly a power law, $\rho\propto r^{-\gamma}$ with
$\gamma=\half(3-b)$. The eigenfrequencies were computed using the Kalnajs
matrix method \citep{pol81,wein91,saha91} using the potential basis functions
$\Phi_p(r)=ru^p/(1+r)^3$, where $u=(r-1)/(r+1)$ and $p$ is an integer
\citep{ho92,saha93}. Typically 20 basis functions were used.

For example, consider a \df\ with $h=0.6$, $b=0$, $I_{\rm min}=1$ and $I_{\rm
max}=(30)^{1/2}$ (in units where $GM=1$). There is a gap in the continuous
spectrum given by $0<\omega^2<\Omega_{\rm min}^2=14.68$, in which there is an
isolated mode at $\omega^2=8.0$ to within 3\%.

\subsection{Stability for $l>1$}

\label{sec:lbig}

\noindent
We have no analytic results on stability for $l>1$. However, we
have searched numerically for unstable modes with $l=2$ in stellar systems
with the \df\ (\ref{eq:dfdefa}), using Goodman's (1988) sufficient criterion
for instability.  In the current context and notation, this condition can be
stated as: a spherical near-Keplerian stellar system is secularly unstable if
there is a trial potential and density $\Phi_{lm}(r)$ and $\rho_{lm}(r)$,
satisfying Poisson's equation (\ref{eq:poisint}), such that
\be
-{2^5\pi^3\over 2l+1}\sum_{j=1}^l y_{lj}^2\int {dI_1I_2dI_2\over\Omega_2}
{\p F\over \p I_2}|\langle \Phi_{lm}(r)\cos j\psi\rangle|^2 > - \int dr\, r^2
\Phi_{lm}^\ast \rho_{lm}.
\label{eq:goodman}
\ee
This result does not require any restrictions on the \df\ $F(I_1,I_2)$.

We have explored a variety of trial functions and several values of the
parameters $I_{\rm max}/I_{\rm min}$, $b$, and $h$ that define the \df. While
the search was not exhaustive, we found no evidence of instability.

\subsection{Secular modes in flat systems}

\label{sec:flat}

\noindent
Zero-thickness near-Keplerian disks provide an instructive contrast to
spherical systems, and also provide some indication of how the behavior of
spherical systems is likely to be modified by flattening. Flat systems are
less simple than spherical systems, for two main reasons: (1) the apsidal
precession rate $\Omega_2$ is not necessarily negative, as it is for spherical
systems (\S\ref{sec:first}); thus there is no simple definition of an inner
product like equation (\ref{eq:inner}) that makes the operator $\vS^l$
Hermitian; (2) the relation between the mass distribution $M_\ast(R)$ and the
potential $\Phi_\ast(R)$ is more complicated for flat systems than for
spherical ones. 

We assume that motion is restricted to a plane described by polar coordinates
$(R,\phi)$. We may continue to use the actions $I_1$ and $I_2$ and the
conjugate angles $w_1$ and $w_2=\omega$ (eqs. \ref{eq:acdef} and
\ref{eq:angledef}), except that now the angular momentum $I_2$ is a signed
variable ($I_2>0$ for prograde orbits and $I_2<0$ for retrograde orbits), and
the argument of pericenter $w_2$ is measured from the zero of azimuth in the
direction of increasing azimuth (instead of from the ascending node in the
direction of orbital motion). The true anomaly $\psi$ is measured from
pericenter in the direction of increasing azimuth rather than the direction of
orbital motion; thus the azimuthal angle $\phi=w_2+\psi$. With these
definitions, equation (\ref{eq:pomegadot}) becomes
\be
\Omega_2={I_2\over GMe}\left\langle{M_\ast(r)\cos
\psi\over r^2}\right\rangle,
\label{eq:revpomega}
\ee
where both $\Omega_2$ and $I_2$ may be positive or negative. 

According to Jeans's theorem, the equilibrium \df\ may be written
$F(I_1,I_2)$; since we are interested in analogs to spherical systems we shall
assume that the \df\ is an even function of the velocity, or an even function
of $I_2$. Most previous studies of secular stability of near-Keplerian disks,
which date back to the work of Laplace and Lagrange on the stability of the
solar system \citep{mur99}, have focused on systems in which all stars
orbit in the same direction \citep{sri99,tre01}.

We restrict ourselves to lopsided potential perturbations, of the form 
\be
\Phi_p(\vx)=\Phi_1(R)\cos\phi.
\label{eq:potpert}
\ee
In the secular approximation, we may average this potential over the fast angle
$w_1$, to obtain
\be
\Phi_s(\vx)=\langle\Phi_1(R)\cos\psi\rangle\cos\omega
\ee
(compare eq.\ [\ref{eq:phisec}]). Following the arguments in equations
(\ref{eq:hamsec})--(\ref{eq:dfpert}), the response \df\ is 
\be
f_r={\p F\over\p I_2}{\langle\Phi_1(R)\cos\psi\rangle\over\Omega_2}\cos\omega.
\label{eq:dfff}
\ee
The response surface density is $\mu_r(\vx)=\int d\vv f_r$. To evaluate this
we use the relation, derivable from equation (\ref{eq:evec}),
\be
e\cos\omega=\cos\phi\left({Rv_\phi^2\over GM}-1\right)+{Rs_Rv_Rv_\phi\over
GM}\sin\phi; 
\label{eq:ecosom}
\ee 
as usual $v_R=(2E+2GM/R-I_2^2/R^2)^{1/2}>0$ is the radial speed and $s_R=\pm
1$ is the sign of the radial velocity; and $v_\phi$ is the azimuthal velocity,
which can be positive or negative. The contributions of the second
term in equation (\ref{eq:ecosom}) to $\mu_r(\vx)$ from $s_R=\pm 1$ average to
zero. Thus
\be
\mu_r(\vx)=2\cos\phi\int_0^\infty dv_R \int_{-\infty}^\infty dv_\phi
{\p F\over\p I_2}{\langle\Phi_1(R)\cos\psi\rangle\over e\Omega_2}
\left({Rv_\phi^2\over GM}-1\right).
\label{eq:denint}
\ee
We convert this into an integral over action space using $I_2=Rv_\phi$ and
equation (\ref{eq:didvr}), 
\be
\mu_r(\vx)={2\cos\phi\over R}\int_0^\infty dI_1\Omega_1\int_{-I_1}^{I_1} 
{dI_2\over e\Omega_2v_R}{\p F\over\p I_2}\langle\Phi_1(R)\cos\psi\rangle
\left({I_2^2\over GMR}-1\right).
\label{eq:deninta}
\ee

Now consider the case in which the potential perturbation (\ref{eq:potpert})
corresponds to a uniform displacement of the unperturbed stellar potential
$\Phi_\ast(\vx)$ by an amount $\xi$ in the $x$-direction. Then
\be
\Phi_p(\vx)=-\xi\cos\phi {d\Phi_\ast\over dR}=\Phi_1(R)\cos\phi
\qquad\hbox{where}\qquad \Phi_1(R)=-\xi{d\Phi_\ast\over dR}.
\ee
The corresponding surface-density perturbation is
\be
\mu_p(\vx)=-\xi\cos\phi {d\mu_\ast\over dR}, 
\label{eq:mup}
\ee
where $\mu_\ast(R)$ is the surface density of the unperturbed
disk. Substituting this form for $\Phi_1(R)$ into equation (\ref{eq:deninta})
and using equation (\ref{eq:revpomega}) to simplify the result,
\be
\mu_r(\vx)=-{2GM\xi\cos\phi\over R}\int_0^\infty dI_1\Omega_1\int_{-I_1}^{I_1}
{dI_2\over I_2v_R}{\p F\over\p I_2}\left({I_2^2\over GMR}-1\right).
\label{eq:denintb}
\ee
Using equation (\ref{eq:partial}) this simplifies further to
\be
\mu_r(\vx)=-2\xi R\cos\phi {\p\over\p R}\int_0^\infty
dI_1\Omega_1\int_{-I_1}^{I_1}{ dI_2 v_R\over I_2}
{\p F\over\p I_2}.
\label{eq:denintc}
\ee

For comparison, the density perturbation (\ref{eq:mup}) that gives rise to the
perturbing potential is
\be
\mu_p(\vx)=-\xi\cos\phi{\p\over\p R}\int d\vv F=
-2\xi\cos\phi{\p\over\p R}{1\over R}\int {dI_1 dI_2 \Omega_1\over
v_R} F;
\label{eq:denintd}
\ee
this is not the same as the response density $\mu_r(\vx)$ of equation
(\ref{eq:denintc}) and hence a neutral mode with this density distribution
does not generally exist in flat near-Keplerian systems. 

The densities that we have derived are useful as trial functions to investigate
the secular stability of such systems using Goodman's (1988) criterion. 
For disk systems Goodman's criterion states that if there exists a trial
surface-density or potential perturbation $\mu_p(\vx)$ or $\Phi_p(\vx)$ with
time dependence $\exp(\lambda t)$, $\lambda>0$, such that the response density
$\mu_r(\vx)$ satisfies
\be
\int d\vx \Phi_p(\vx)[\mu_p(\vx)-\mu_r(\vx)]>0,
\ee
then the stellar system is unstable. This result requires only that the \df\
is an even function of the velocities.

In our case we let $\lambda\to 0$ and use
equations (\ref{eq:denintc}) and (\ref{eq:denintd}); thus
\be
\int d\vx \Phi_p(\vx)[\mu_p(\vx)-\mu_r(\vx)]=2\pi\xi^2\int
dR\,R{d\Phi_\ast\over dR}\int_0^\infty dI_1\Omega_1\int_{-I_1}^{I_1} dI_2\left(
F{\p\over\p R}{1\over Rv_R} - {1\over I_2}{\p F\over\p I_2}R{\p v_R\over\p R}
\right). 
\ee
We now integrate the term involving $\p F/\p I_2$ by parts, assuming that the
loss cone is empty (more precisely, we assume that $F/I_2$ is non-singular as
$I_2\to 0$) so that there is no divergence at $I_2=0$; the other boundary
term vanishes since the upper limit to the range of integration is set by
$v_R=0$. Using the relations (\ref{eq:partial}) and (\ref{eq:ident}) we obtain
\be
\int d\vx \Phi_p(\vx)[\mu_p(\vx)-\mu_r(\vx)]=2\pi GM\xi^2\int
dR\,{d\Phi_\ast\over dR}\int_0^\infty dI_1\Omega_1\int_{-I_1}^{I_1}
{dI_2\over v_R I_2^2}F.
\label{eq:flatstable}
\ee
If this expression is positive then the stellar system is secularly
unstable (this result does not require either that $\p F/\p |I_2|>0$ or
that $\Omega_2<0$). The inner integral is positive, but in flat systems
$d\Phi_\ast/dR$ is not necessarily positive everywhere (in particular, disks
with zero surface density near the center, as one expects if the loss cone is
empty, have $\Phi_\ast(R)=\Phi_0-\half \alpha R^2 + \hbox{O}(R^4)$ near the
center, with $\alpha >0$). Nevertheless, many disks have $d\Phi_\ast/dR>0$
throughout the radial range containing most of the disk mass, and thus are
likely to be secularly unstable.

It has long been known that disk systems with counter-rotating stars are prone
to lopsided instabilities, but these analyses have focused on isolated stellar
disks or disks embedded in a massive halo \citep{zang78,ara87,pp90,sm94},
rather than near-Keplerian disks \citep{tou02}. The results from
this Section imply that lopsided instabilities are also common in
near-Keplerian disks with little or no net rotation.

\section{Summary}

\label{sec:summary}

\noindent
Near-Keplerian stellar systems can support secular or slow normal modes, in
which the eigenfrequency is of order $t_{\rm dyn}^{-1}(M_\ast/M)$, where
$t_{\rm dyn}$ is the dynamical time, $M$ is the mass of the central object,
and $M_\ast$ is the mass of the stellar system.  We have shown that spherical
near-Keplerian systems in which the \df\ is a decreasing function of angular
momentum, or an increasing function of angular momentum that is non-zero at
zero angular momentum, are stable to $l=1$ secular modes. Numerical
calculations suggest that these systems are also generally stable to modes with
$l>1$.

Spherical near-Keplerian systems in which the \df\ is an increasing function
of angular momentum and zero at zero angular momentum (an empty loss cone)
also have no unstable modes but, remarkably, are all neutrally stable to an
$l=1$ mode. The spatial form of the neutral mode corresponds to a uniform
displacement of the stellar system relative to the central mass, although the
velocity-space perturbation does not. The analogous
zero-thickness disks are often unstable to $l=1$ modes.

Other authors have investigated lopsided modes of stellar systems, but in
quite different contexts from the present paper. \cite{wein94} has reported
nearly neutral $l=1$ ``sloshing'' modes in linear stability analyses of
spherical stellar systems, and long-lived oscillations in the centers of
$N$-body models have been reported by several authors
\citep{ms92,sa96}. However, these systems do not contain a central massive
object so the mechanism responsible for the oscillations is probably quite
different. \cite{ti98} have observed oscillations of a massive central object
in $N$-body simulations; however, they argue that rotation of the stellar
system is necessary to drive the oscillation, so again the cause of the 
oscillations is likely to be different. 

Our results suggest that the stellar systems found in the centers of galaxies
containing a black hole are susceptible to lopsided distortions. 
Of course, the existence of neutral modes does not imply instability, only
that the system is close to instability in some sense. Our analysis has
ignored a number of smaller effects, which may modify the neutral mode to one
that is weakly stable or unstable. These include:

\begin{itemize}

\item Relativistic effects. These induce apsidal precession at a rate
\be
\dot\omega_{\rm GR}={3(GM)^{3/2}\over a^{5/2}c^2(1-e^2)}.
\label{eq:gr}
\ee
Since relativistic precession is prograde, while precession due to
self-gravity is retrograde, small relativistic corrections increase
$-1/\Omega_2$ and hence enhance instability according to Goodman's criterion
(\ref{eq:goodman}). On the other hand, once the relativistic precession rate
exceeds the precession rate due to self-gravity it promotes secular
stability. Precession due to the giant planets is also prograde, so has
similar effects on the Oort cloud. 

\item Flattening. Flattened, non-rotating systems are likely to be less stable
than spherical ones, since zero-thickness systems are often unstable
(\S\ref{sec:flat}). 

\item Rotation. This probably promotes stability, since
rotating near-Keplerian disk systems of stars on nearly circular orbits are
stable \citep{tre01}.

\item Non-resonant contributions to the eigenvalue equation. These are smaller
by of order $M_\ast/M$ and in the usual case when $\p F/\p I_1<0$ will promote
stability. 

\end{itemize}

A natural next step is to examine the secular stability of near-Keplerian
stellar systems using numerical experiments. This is an arena in which there
has been surprisingly little activity. Most investigations of near-Keplerian
systems focus on their long-term evolution, using the Fokker-Planck
approximation and assuming spherical symmetry
\citep{bw76,ck78,fre01}. Recent $N$-body simulations of near-Keplerian systems
do not have sufficient resolution to probe the region in which the loss
cone is empty \citep{baum04,pre04}. 

A related question is whether we can extend the linearized analysis of this
paper to construct self-consistent models of near-Keplerian systems with
significant lopsidedness. The general properties of orbits in lopsided
near-Keplerian potentials were discussed by \cite{st99} but they did not
construct self-consistent models. 

If near-Keplerian stellar systems are secularly unstable, then standard models
of the distribution of stars around black holes and estimates of the rate at
which black holes consume or disrupt stars may be quite misleading. 

\acknowledgements

This research was supported in part by NASA grants NNG04H44g and NNG04GL47G. I
thank Eric Ford for his contributions to this work at an early stage, and
Agris Kalnajs and Alar Toomre for thoughtful comments on the manuscript.

\end{document}